\documentclass[twocolumn,traditabstract]{aa}
\usepackage{txfonts}
\usepackage{graphicx}
\usepackage{natbib}
\renewcommand{\deg}{\mbox{$^{\circ}$}~}
\newcommand{\kms}{\mbox{km~s$^{-1}$}}
\newcommand{\mols}{\mbox{molec.~s$^{-1}$}}

\begin{document}

\title{The extraordinary composition of the blue comet C/2016~R2 (PanSTARRS)
\thanks{Based on observations carried out with the IRAM 30m telescope.
IRAM is supported by INSU/CNRS (France), MPG (Germany) and IGN (Spain).}
\thanks{The radio spectra are available at the CDS via anonymous 
ftp to cdsarc.u-strasbg.fr (130.79.128.5)
or via http://cdsweb.u-strasbg.fr/cgi-bin/qcat?J/A+A/}
}
 
\author{N. Biver\inst{1,5}
   \and D. Bockel\'ee-Morvan\inst{1}
   \and G. Paubert\inst{2}
   \and R. Moreno\inst{1}
   \and J. Crovisier\inst{1}
   \and J. Boissier\inst{3}
   \and E. Bertrand\inst{4,5}
   \and H. Boussier\inst{4,5}
   \and F. Kugel\inst{6,5}
   \and A. McKay\inst{7}
   \and N. Dello Russo\inst{8}
   \and M. A. DiSanti\inst{9}
   }

\institute{LESIA, Observatoire de Paris, PSL Research University, CNRS, 
Sorbonne Universit\'e, Univ. Paris Diderot, 
Sorbonne Paris Cit\'e, 5 place Jules Janssen, F-92195 Meudon, France
 \and IRAM, Avd. Divina Pastora, 7, 18012 Granada, Spain
 \and IRAM, 300, rue de la Piscine, F-38406 Saint Martin d'H\`eres, France
 \and Astronomical Ring for Access to Spectroscopy (ARAS) (http://www.astrosurf.com/aras/intro/intro.htm)
 \and Commission des com\`etes, Soci\'et\'e Astronomique de France, 3 rue Beethoven, F-75016 Paris, France
 \and Observatoire de Dauban, 04 Banon, France
\and NASA GSFC/USRA, 8800 Greenbelt Rd, Greenbelt, MD 20771, USA 
\and Johns Hopkins University Applied Physics Laboratory, 11100 Johns Hopkins Rd., Laurel, MD 20723, USA
\and NASA Goddard Center for Astrobiology, NASA GSFC, Mail Stop 690, Greenbelt, MD 20771, USA
}

   \titlerunning{Composition of comet C/2016~R2 (PanSTARRS)}
   \authorrunning{Biver et al.}
   \date{\today}

   \abstract{We present a multi-wavelength study of comet C/2016~R2 (PanSTARRS).
This comet was observed on 23-24 January 2018 with the IRAM~30m telescope, and
in January to March 2018 with the Nan\c{c}ay radio telescope. Visible
spectroscopy was performed in December 2017 and February 2018 with small
amateur telescopes. We report on measurements of CO, CH$_3$OH, H$_2$CO and
HCN production rates, and on the determination of the N$_2$/CO abundance ratio.
Several other species, especially OH, were searched for but not detected.
The inferred relative abundances, including upper limits for sulfur
species, are compared to those measured in other comets at about the same
heliocentric distance of $\sim$2.8 AU. The coma composition of comet C/2016~R2
is very different from all other comets observed so far, being
 rich in N$_2$ and CO and dust poor. This suggests that this
comet might belong to a very rare group of comets formed beyond the N$_2$
ice line. Alternatively, comet C/2016~R2 (PanSTARRS) could be the fragment of
a large and differentiated transneptunian object, with properties
characteristic of volatile-enriched layers. } 


\maketitle

\section{Introduction}
Comets are the most pristine remnants of the formation of the
solar system 4.6 billion years ago. Investigating the
composition of cometary ices provides clues to the physical
conditions and chemical processes at play in the primitive solar
nebula. Comets may also have played a role in the delivery of
water and organic material to the early Earth 
\citep[see][and references therein]{Har11}.
The latest simulations of the evolution of the early solar system \citep{Bra13,Obr14}
suggest a more complex scenario. On the one hand, ice-rich bodies formed beyond
Jupiter may have been implanted in the outer asteroid belt, participating in
the supply of water to the Earth, or, on the other hand, current comets coming
from either the Oort Cloud or the scattered disk of the Kuiper belt may have
formed in the same trans-Neptunian region, sampling the same diversity of 
formation conditions. Understanding the diversity in composition and isotopic
ratios of the comet material is therefore essential for the assessment of such
scenarios \citep{Alt03,Boc15}.  

Recent years have seen significant improvement in the sensitivity
and spectral coverage of millimetre receivers enabling sensitive
spectral surveys of cometary atmospheres and simultaneous 
observations of several molecules.
We report here observations of a very peculiar comet, C/2016~R2 (PanSTARRS),
with the 30-m telescope of the Institut de radioastronomie
millim\'etrique (IRAM). This object is a long-period, dynamically
old, Oort cloud comet that passed close to the Sun 21\,600 years ago
and has an orbit inclination of 58\degr~.
It is very peculiar in the sense that in autumn 2017, while approaching
the Sun (perihelion was on 9.6
May 2018 UT at 2.60 AU), it exhibited a
deep blue coma and tail, due to the presence of strong CO$^+$ lines in the
optical spectrum and very little dust or other emission lines.
In contrast, usual comets show a dust tail and a coma of
neutral or yellowish colour, resulting from the scattering of solar
radiation by dust and the emission of C$_2$ Swan bands, if any at this
heliocentric distance, which are not seen here.
It seems to belong to a category of comets of which we know only 
very few examples: C/1908~R1 (Morehouse) \citep{Bau11} or C/1961~R1 (Humason)
\citep{Gre62}, none having been observed with modern astronomical facilities.
As in those two comets that were however observed closer to the Sun, optical
spectroscopy of comet C/2016~R2 revealed unusually strong N$_2^+$ lines
\citep{Coc18}.

Here we present millimetre observations of CO, CH$_3$OH, H$_2$CO, HCN, CO$^+$
and upper limits on several other molecules obtained with the IRAM~30-m,
complemented by observations of the OH radical with the Nan\c{c}ay
radio telescope, and amateur observations of CO$^+$, N$_2^+$ and dust continuum.
From the millimetre data
we derive constraints on the gas temperature and outgassing pattern. Production
rates and their time evolution on 23--24 January, as well as abundances
relative to CH$_3$OH or CO are derived.
We then compare the abundances in comet C/2016~R2 with those measured in eight
other comets observed at a similar heliocentric distance.


\section{Observations of comet C/2016~R2 (PanSTARRS)}

\subsection{Observations conducted at IRAM~30-m}

Comet C/2016 R2 (PanSTARRS) was observed with the IRAM~30-m radio telescope
on two consecutive evenings, 23.8 and 24.8 January 2018 UT, under good
weather (mean precipitable water vapour of 1~mm).
We targeted the CO($J$=2--1) line at 230538.000~MHz first, before the HCN($J$=3--2) line at 265886.434~MHz and the H$_2$S($1_{10}-1_{01}$) line
at 168762.762~MHz \citep[see CDMS catalogue,][the line frequencies
uncertainties (< 1~kHz) being much lower than the finest sampling of the
spectrometer used (20~kHz)]{CDMS},
with other molecular lines, especially of methanol,
in the band or the other side band. We used the EMIR 1~mm and 2~mm band
receivers \citep{Car12} with the fast Fourier transform
spectrometer (FTS) set to a 200 kHz sampling covering $\sim2\times8$ GHz
in two polarizations simultaneously.

\begin{figure}[ht]
\centering
\resizebox{\hsize}{!}{\includegraphics[angle=270]{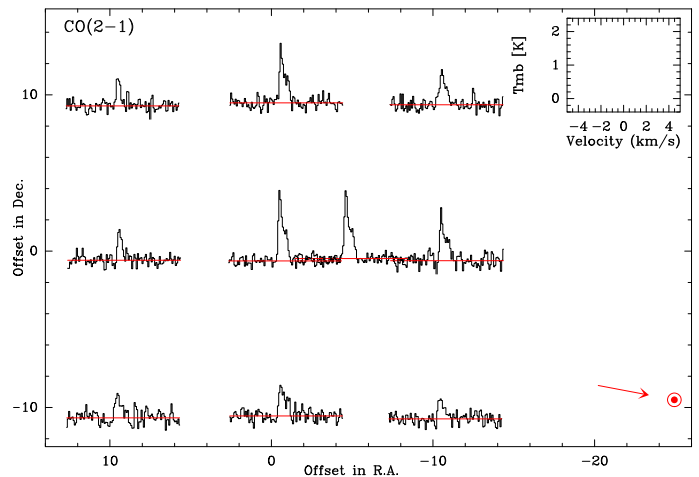}}
\caption{Small map of the CO(2-1) emission in the coma
    of comet C/2016~R2 (PanSTARRS) obtained with the IRAM-30m
    telescope on 24.85 January 2018. The velocity and intensity scales of the
    spectra are given in the top right.
    The arrow points towards 
    the projected direction of the Sun (phase angle of 17\degr).}
\label{figmapco}
\end{figure}

\begin{figure}[ht]
\centering
\resizebox{\hsize}{!}{\includegraphics[angle=270]{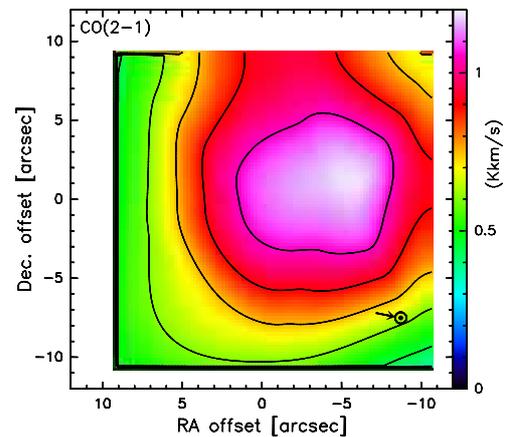}}
\caption{Colour-coded map of the CO(2-1) line integrated intensity
    of comet C/2016~R2 (PanSTARRS) obtained with the IRAM~30-m
    telescope on 24.85 January 2018. The (0,0) offset position corresponds
    to the JPL ephemeris position, with a $\sim2$\arcsec~ pointing uncertainty.
    The arrow points towards 
    the projected direction of the Sun (phase angle of 17\degr).}
\label{figimgco}
\end{figure}

A log of the observations is given in
Table~\ref{tablog}. We used the secondary mirror wobbler, with a 180\arcsec~
throw at a frequency of 0.5~Hz to cancel the sky background.
The comet was tracked with JPL\#14 orbital solution. Coarse
mapping of the strong CO line shows that the peak intensity was 
shifted by 2--3\arcsec~ to the west northwest (WNW) (Figs.~\ref{figmapco} and \ref{figimgco}).
We adjusted the pointing offset to integrate closer to the peak of intensity.
All pointing offsets mentioned henceforth in Tables~\ref{tablog} and
\ref{tabobsoff} are relative to the position of this peak of intensity.

\begin{figure}
\centering
\resizebox{\hsize}{!}{\includegraphics[angle=270]{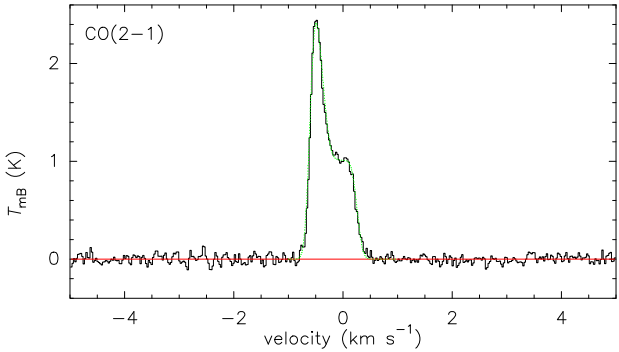}}
\caption{Average on-nucleus (offset 1.7\arcsec) spectrum of the CO(2-1) line
at 230.538~GHz obtained with the IRAM 30m telescope on 24.80 January 2018.
The vertical scale is the main beam brightness temperature
and the horizontal scale is the Doppler velocity in the comet rest frame.
The simulated profile with $Q_{CO} = 5\times10^{28}$~\mols~ in a 0--60\deg
cone at $v_{exp}=0.56$~\kms~ and $Q_{CO} = 5.5\times10^{28}$~\mols~
in a 60--120\deg cone at $v_{exp}=0.50$~\kms~ is superimposed
in green dotted line.}
\label{figcocenter}
\end{figure}

\begin{figure}
\centering
\resizebox{\hsize}{!}{\includegraphics[angle=270]{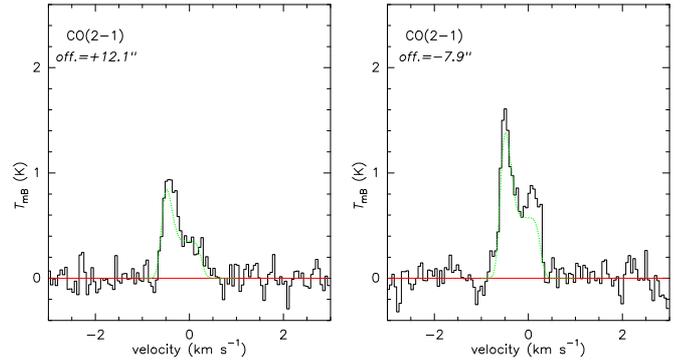}}
\caption{Average off nucleus spectra of the CO(2-1)
line obtained with the IRAM 30m telescope on 23.78--24.85 January 2018:
left: 12.1\arcsec tail-ward ($\Delta$RA$>0$),
right: 7.9\arcsec sun-ward.
Scales are as in Fig.~\ref{figcocenter}.
The simulated profiles with $Q_{CO} = 5\times10^{28}$~\mols~ in a 0--60\deg
cone at $v_{exp}=0.56$~\kms and $Q_{CO} = 5.5\times10^{28}$~\mols~
in a 60--120\deg cone at $v_{exp}=0.50$~\kms are superimposed
in green dotted line.}
\label{figcooffset}
\end{figure}

\begin{figure}
\centering
\resizebox{\hsize}{!}{\includegraphics[angle=270]{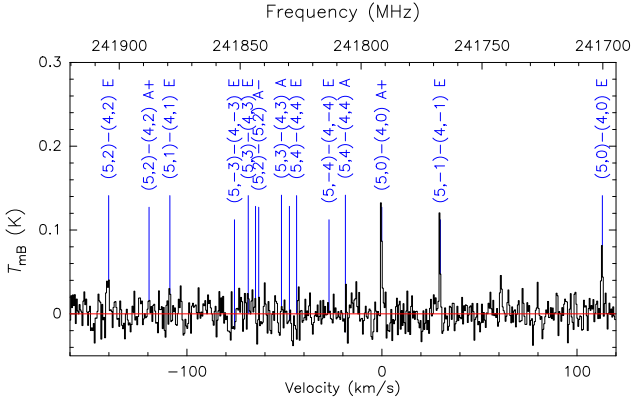}}
\caption{Series of methanol lines around 242~GHz from comet
C/2016~R2 (PanSTARRS) obtained with the IRAM-30m telescope on 23.8 January
2018. The inferred rotational temperature is $T_{rot}({\rm 242GHz})=18\pm2$ K.
The vertical scale is the main beam brightness temperature
and the horizontal scale is the Doppler velocity in the comet rest frame
(lower axis) or the rest frequency (upper axis).}
\label{figmet241}
\end{figure}

\begin{figure}
\centering
\resizebox{\hsize}{!}{\includegraphics[angle=270]{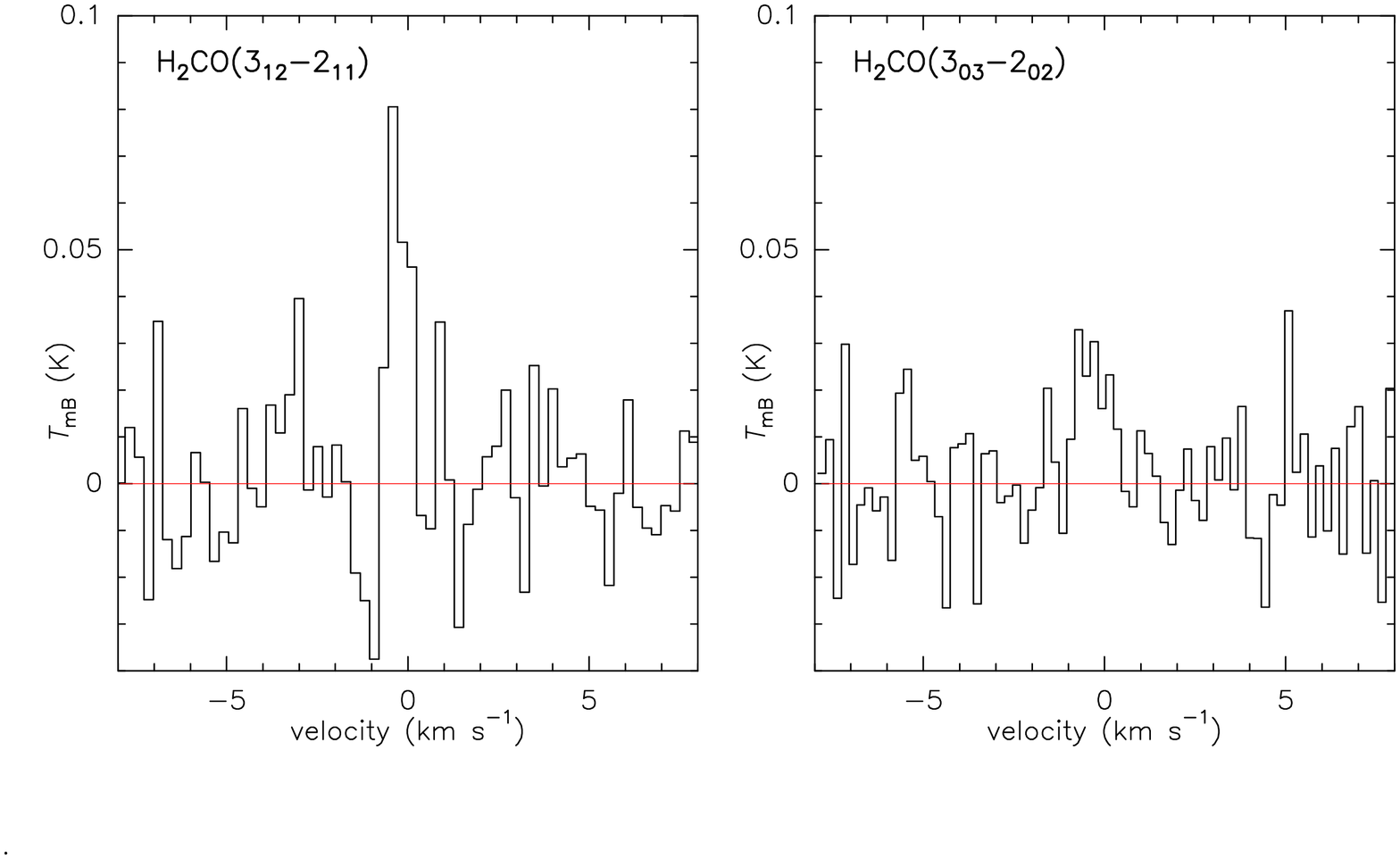}}
\caption{Average on-nucleus spectra of the ortho H$_2$CO($3_{12}-2_{11}$) line
at 225.698~GHz (left) and para H$_2$CO($3_{03}-2_{02}$) line
at 218.222~GHz (right) in comet C/2016 R2 (PanSTARRS)
obtained with the IRAM 30m telescope on 23.81 and 24.78 January 2018,
respectively. Scales as in Fig.~\ref{figcocenter}.}
\label{figh2co}
\end{figure}

\begin{figure}
\centering
\resizebox{\hsize}{!}{\includegraphics[angle=270]{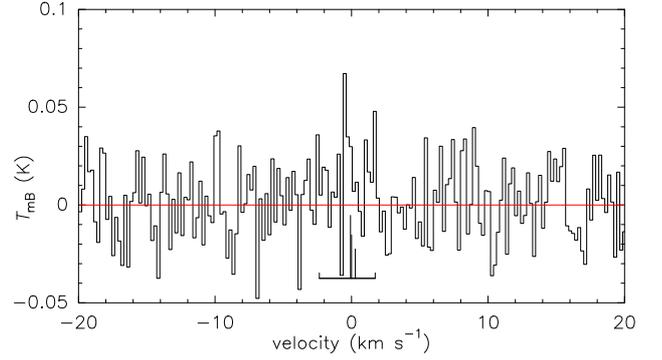}}
\caption{Average on-nucleus spectrum of the HCN~(3-2) 
line at 265886.434~MHz obtained with the IRAM 30m telescope between 23 and
24 January 2018. Scales are as in Fig.~\ref{figcocenter}. Relative intensities
and positions of the hyperfine components relative to the $F=3-2$ line at
265886.434~MHz are indicated.}
\label{fighcn}
\end{figure}

The comet immediately revealed itself as very different from most other comets
observed so far. The CO(2-1) line is one order of magnitude stronger than
expected (based on a correlation between visual magnitudes and CO production
rates from \citet{Biv01}).
Meanwhile the narrow and blue-shifted line shape (Fig.~\ref{figcocenter})
strongly resembles the one of distant comet 29P/Schwassmann-Wachmann 1
\citep{Gun08}, whose activity is dominated by a continuous, asymmetric and
large outgassing of CO ($\approx4\times10^{28}$\mols).
In total, 15 methanol lines are clearly
detected; two lines of formaldehyde are also detected and the HCN(3-2)
line is marginal.
The sum of the three lines of HNCO covered by the observations
that are expected to be the strongest is close to $3\sigma$.
Lines of neutral species (Figs.\ref{figcocenter}--\ref{fighcn})
extend from -0.7 to +0.5\kms on the axis of Doppler velocities
relative to the comet rest frame, which is due
to the projection of the expansion velocity vector of the gas. 
In contrast, the two CO$^+$ lines at 236062.553 and 235789.641~MHz
show a very marginal feature of  $\sim2$~\kms~ in width superimposed
on a more pronounced broad red-shifted line of 30~\kms~.
No other species are detected.

\begin{figure}
\centering
\resizebox{\hsize}{!}{\includegraphics[angle=270]{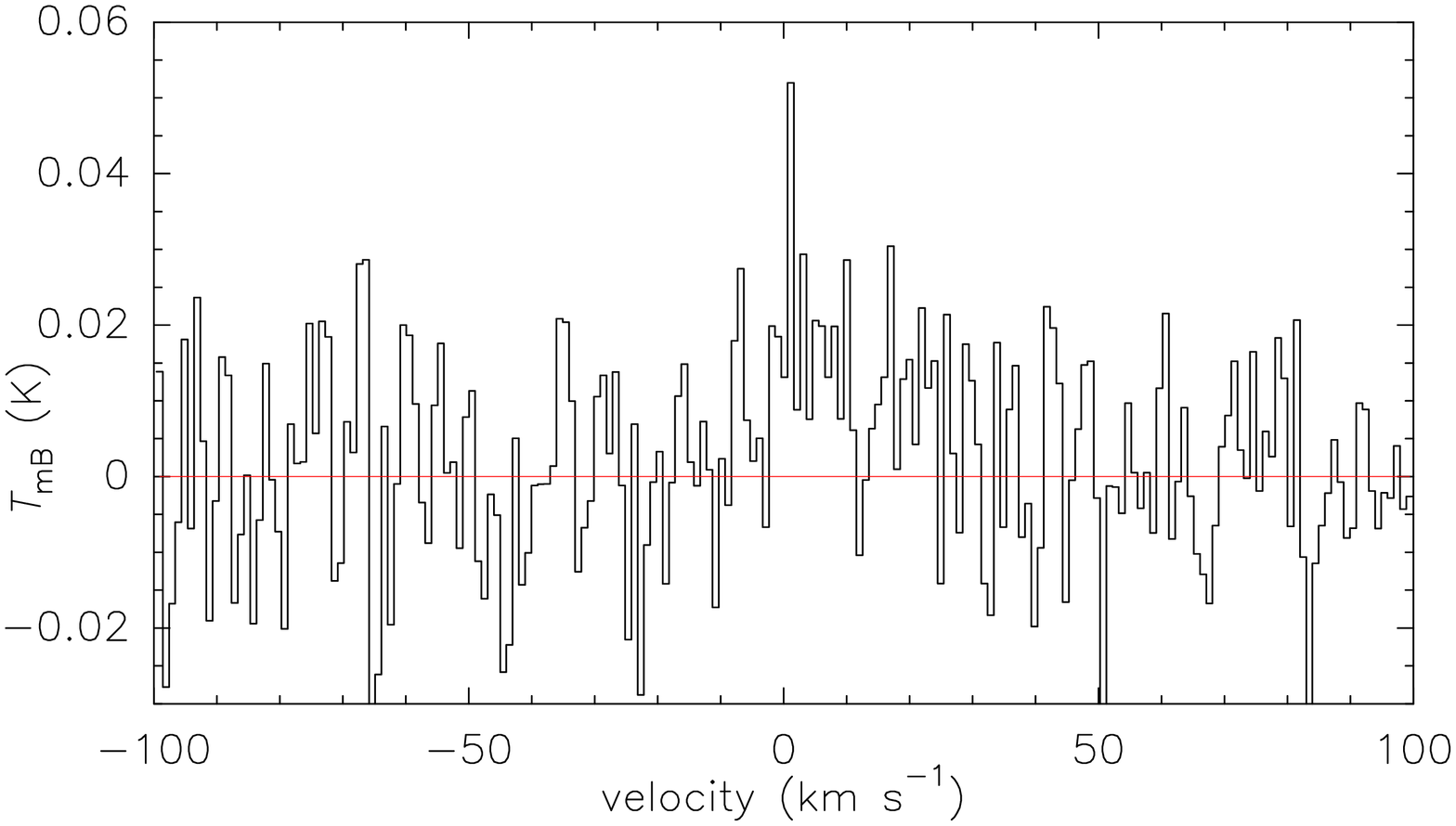}}
\caption{Weighted average of the two strongest CO$^+$ $N=2-1$ lines observed in
C/2016~R2 (PanSTARRS) on 24.8 January 2018. Each line has been divided
by its statistical weight (9/15 and 5/15 for the $F=5/2-3/2$ and
$F=3/2-1/2$ transitions respectively) before averaging.
Scales are as in Fig.~\ref{figcocenter}.
The line shows a narrow peak close to the zero velocity and a broader component
mostly red-shifted indicative of antisunward acceleration of CO$^+$ ions by the
solar wind.}
\label{figcoplus}
\end{figure}

Sample spectra are shown in Figs.~\ref{figmapco}--\ref{fighcn} and
\ref{figcoplus}, and the line intensities and derived production rates
are given in Table~\ref{tabobsqp}.

\begin{table*}
\caption[]{Log of IRAM observations.}\label{tablog}\vspace{-0.2cm}
\begin{center}
\begin{tabular}{rcccccc}
\hline\hline
UT date  & $<r_{h}>$  & $<\Delta>$   & Offset & Integ. time & Freq. range \\
$($yyyy/mm/dd.d--dd.d) & (AU)  & (AU)  & (\arcsec)\tablefootmark{a} & (min)\tablefootmark{b} & (GHz)  \\
\hline
2018/01/23.78--23.84 & 2.832 & 2.214 & 3.0 & 61.0 & 224.9--232.7, 240.5--248.3 \\
      23.87--23.95 & 2.832 & 2.216 & 2.3 & 74.7  & 248.7--256.5, 264.4--272.2 \\
      23.98--24.04 & 2.831 & 2.217 & 3.9 & 56.0  & 146.9--154.7, 162.6--170.4 \\
2018/01/24.71--24.86 & 2.828 & 2.224 & 2.7 & 147. & 212.8--220.5, 228.4--236.2 \\
      24.88--24.91 & 2.828 & 2.225 & 1.9 & 28.0  & 248.7--256.5, 264.4--272.2 \\
\hline
\end{tabular}
\end{center}
\tablefoot{
\tablefoottext{a}{Residual pointing offset (relative to the position of peak intensity measured on coarse maps) for the ``On'' nucleus pointings.}
\tablefoottext{b}{``On'' nucleus pointings. But the first tuning of each day includes
15~and 22~min (23.8 and 24.8 January, respectively) spent at 5 to 14\arcsec~
pointing offsets when coarse mapping was carried out.}
}
\end{table*}

\begin{table*}
\caption[]{Observations of OH 18~cm lines at Nan\c{c}ay.}\label{tablognancay}\vspace{-0.2cm}
\begin{center}
\begin{tabular}{lcccccc}
\hline\hline
UT dates  & $<r_{h}>$  & $<\Delta>$   & Mean maser & Line area & $Q_{\rm OH}$ \\
$($yyyy/mm/dd.d--mm/dd.d) & (AU) & (AU) & inversion & (mJy~\kms) & (\mols)\\
\hline
2018/01/02.9--02/09.8 & 2.84 & 2.21 & -0.31 & $-10\pm4$ & $<1.1\times10^{28}$ \\
2018/02/10.8--03/26.7 & 2.69 & 2.68 & -0.23 &  $+6\pm4$ & $<1.7\times10^{28}$ \\
\hline
\end{tabular}
\end{center}
\end{table*}

\subsection{Observation of OH radical with the Nan\c{c}ay radio telescope}
In order to assess the water production rate of the comet, the OH lines
at 1665 and 1667~MHz were searched for with the Nan\c{c}ay radio telescope
between 2 January and 31 March 2018 (observation circumstances
for selected periods are presented in Table~\ref{tablognancay}).
The telescope tracked the comet around transit for about 1~h on average
every two days.

\subsection{Complementary optical observations}
In support of these observations conducted with large facilities, amateur
astronomers followed the activity of the comet (which exhibited a total
visual magnitude between 10 and 11) with their own telescopes. We report here on
results obtained by Etienne Bertrand (St Sordelin), Hubert Boussier
(IAU station K21) and Fran\c{c}ois Kugel (IAU station A77).
The first two astronomers obtained moderate-resolution
($\lambda/\delta\lambda \sim 500$) visible spectra ($\lambda$=390-600~nm)
of the comet, while the latter two  also provided $Af\rho$ values
of the comet (measured outside the range of CO$^+$ lines).
The data and logs of the observations
are also accessible from the comet observation database 
hosted by the Paris Observatory \citep{dbcomet}.

\begin{figure*}[ht]
\centering
\resizebox{\hsize}{!}{\includegraphics[angle=270]{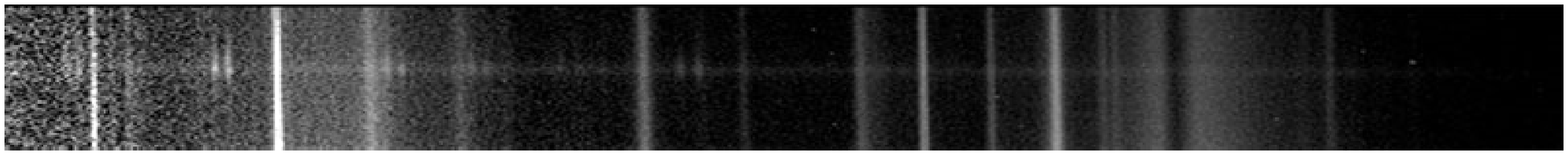}}
\caption{The visible spectrum of comet C/2016~R2 (PanSTARRS) obtained
  by H. Boussier on 22.93 December 2018 UT with a 0.28~m telescope + LISA
  spectrometer and Atik 414L.
  The wavelengths covered are from 385 to 655~nm from left to right. Vertical
  lines are atmospheric lines, mostly due to light pollution, while the
  cometary lines (mostly CO$^+$ doublets) are less extended vertically.
  Scattering of the light of the mercury vapour lamps (Hg-I lines at
    404.7, 435.8, 546.1 and 577+579~nm) and high pressure sodium lamps (568~nm
    line and broad emission around 589~nm) are the dominating sky emissions.}
\label{imgspec-hboussier}
\end{figure*}

\begin{figure*}[ht]
\centering
\resizebox{\hsize}{!}{\includegraphics[angle=270]{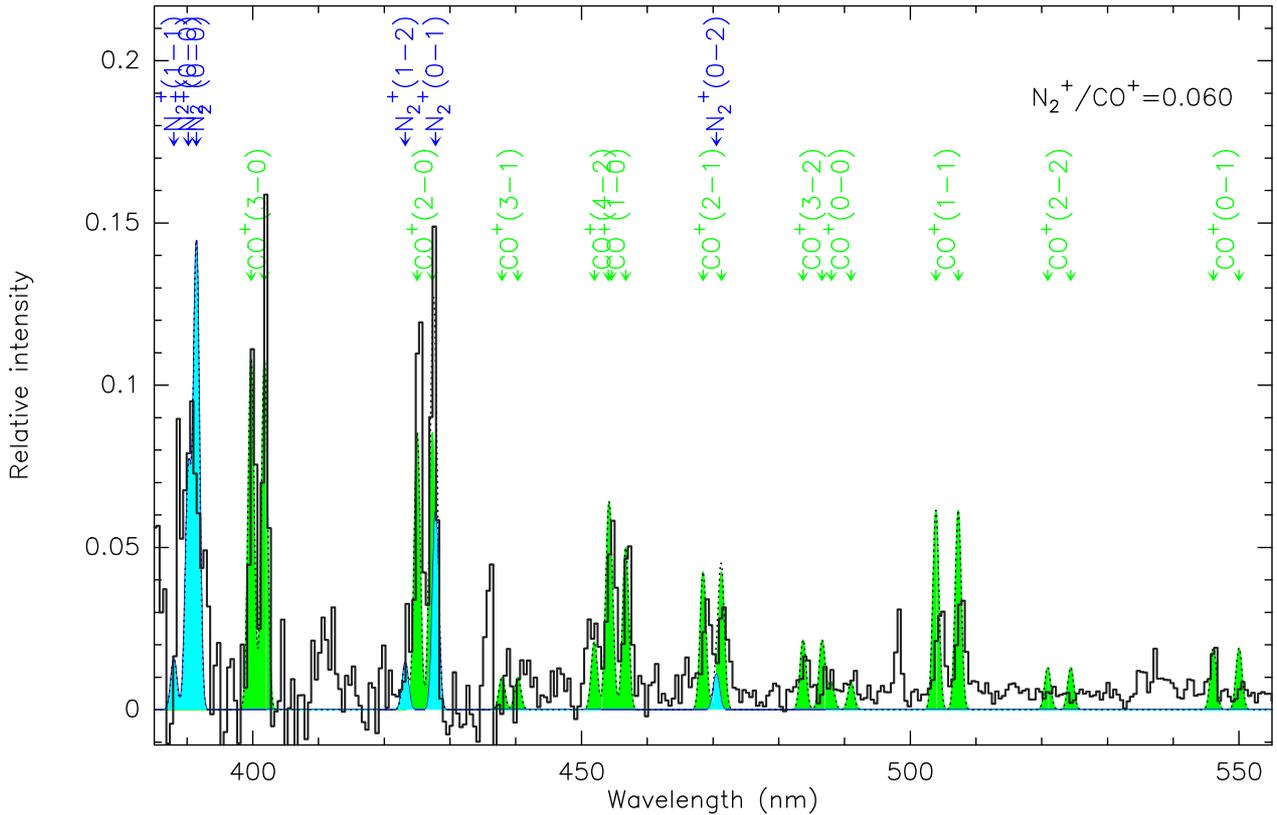}}\vspace{-0.7cm}
\caption{Black line: Visible spectrum of comet C/2016~R2 (PanSTARRS) obtained by
  H. Boussier on 22.92 December 2017 UT (40~min integration) from MPC station
  K21. The spectrum was extracted from the central 10 (binned) pixels rows.
  Green filled lines represent the simulated CO$^+$ spectrum with respective average
  g-factors. Blue filled lines show the N$_2^+$ spectrum for a N$_2^+$/CO$^+$ column
  density ratio of 0.06. The vertical intensity scale has been normalized.
  The dotted line is the sum of N$_2^+$ and CO$^+$ synthetic spectra.
  The signal towards the red end (>500nm) of the spectrum is likely
  under-corrected because the LISA spectrometer is not optimised to cover the
  full wavelength range and the comet was slightly out of focus in that
  part of the spectrum.
  As a consequence, the signal of the lines beyond $\sim$510~nm is lower
  than expected and some lines are not detected. This part of the spectrum
  is not used in the analysis.
}
\label{spec-hboussier}
\end{figure*}

Hubert Boussier took spectra on 29.9 November, and 18.9 and 22.9 December 2017.
In this paper we present an analysis of the spectrum from the latter of the
three (Figs.~\ref{imgspec-hboussier},\ref{spec-hboussier})
obtained between 22:01 and 22:42 UT with a 28-cm Schmidt-cassegrain telescope
working at F/D=7.0 equipped with a LISA spectrometer \citep{shelyak}
and Atik 314L CCD detector.
The slit width and integration window was $2.4\times14$\arcsec.
Sky-glow and light pollution lines were removed using the signal in
$2.4\times33$\arcsec~ windows at -38 and +72\arcsec~ along the slit.
The effective spectrum resolution is 1~nm (5.1~\AA~ per pixel).
Etienne Bertrand obtained his spectrum (Fig.~\ref{spec-hboussier})  on 22 February
2018 between 19:09 and 20:09 UT during a one-hour integration with a 20-cm
Schmidt-cassegrain telescope working at F/D=6.3 equipped with an Alpy600
spectrometer \citep{shelyak} and Atik 414EX CCD detector.
The slit width and integration window was
$3.8\times53$\arcsec. Sky-glow and light pollution lines are removed using
the signal in $3.8\times37$\arcsec~ windows at $\pm177$\arcsec~ along the slit.
The spectral resolution is also
$\lambda/\delta\lambda\sim500$, that is, on the order of 1~nm.
The average apertures used for these observations (equivalent diameter
of 8\arcsec~ and 14\arcsec~ for the 22 Dececember and 22 February
observations, respectively)
are similar to the IRAM~30-m beam, which is useful for comparing both types
of observation.

\begin{figure*}[ht]
\centering
\resizebox{\hsize}{!}{\includegraphics[angle=270]{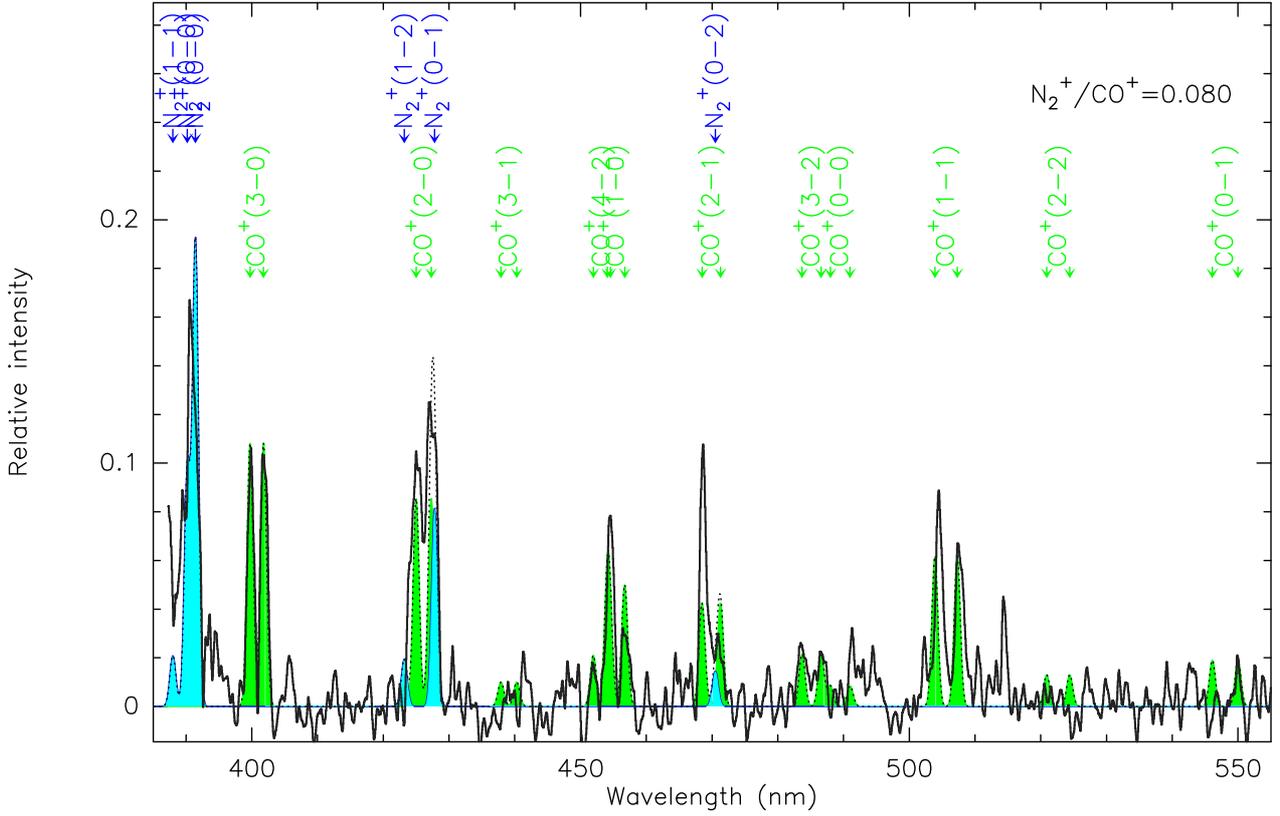}}\vspace{-0.7cm}
\caption{Black line: Visible spectrum of comet C/2016~R2 (PanSTARRS) obtained by
  E. Bertrand on 22.80 February 2018 UT (1~h integration) from Vaux-sur-Mer
  (France). The  spectrum was extracted from the central 50 pixels rows.
  Green filled lines show the simulated CO$^+$ spectrum with respective average
  g-factors. Blue filled lines indicate the N$_2^+$ spectrum for a N$_2^+$/CO$^+$ column
  density ratio of 0.08. The vertical intensity scale has been normalized.
  Dotted lines is the sum of N$_2^+$ and CO$^+$ synthetic spectra.
  }
\label{spec-ebertrand}
\end{figure*}

The optical spectra are presented in 
Figs.~\ref{spec-hboussier} and \ref{spec-ebertrand}.
The correction for the system and sky transmission was done using a
nearby A-type reference star and an out-of-atmosphere synthetic spectrum.
Only a relative calibration was obtained (no absolute calibration of the
fluxes in e.g. Wm$^{-2}$\AA$^{-1}$). Due to poor transmission of the atmosphere
and optical system; and limited sensitivity of the detector towards the blue end
of the spectrum, especially for the LISA spectrometer, the noise and
uncertainty both increase below 400--430~nm. The wavelength calibration was
performed using a reference internal source.

The spectra are clearly dominated by the CO$^+$ doublets contrarily to
most optical spectra of comets, dominated by CN and 
C$_2$ Swan lines \citep[e.g. Fig.1 of][]{Fel04}.
The N$_2^+$ line at 391~nm is also detected in both spectra, but in the noisiest
part of the spectrum; we assumed that there is no contribution from the
nearby CN line at 388~nm, both because it was not detected by \citet{Coc18}
and because the abundance of HCN is very low (see following section).

\section{Analysis and results}
\subsection{IRAM data}
    Thanks to the simultaneous detection of two to six methanol lines and to the
very high S/N obtained on the CO line, we were able to 
constrain the excitation conditions of the molecules in the coma and the
outgassing pattern of the comet.
This is important to derive accurate outgassing rates and relative abundances.

\subsubsection{Gas temperature}
\begin{figure}[ht]
\centering
\resizebox{\hsize}{!}{\includegraphics[angle=270]{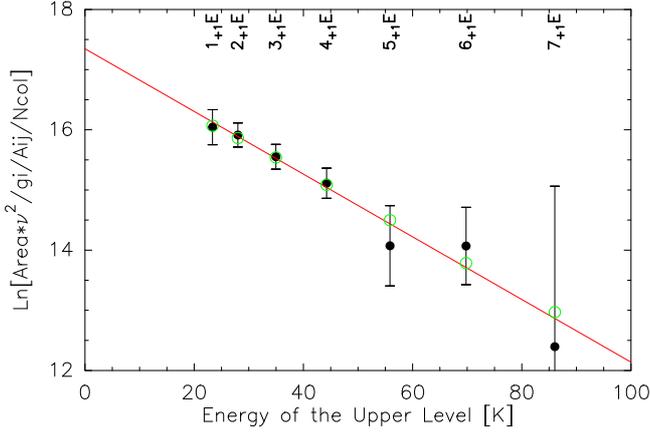}}
\caption{Rotational diagram of the CH$_3$OH lines at 165~GHz.
The green open dots correspond to the expected values for $T_{gas}=23$ K,
corresponding to $T_{rot}=20.5$ K. The red line is the fit to the observations
($T_{rot}=19.2\pm3.9$ K).}
\label{figdiagrot165}
\end{figure}

        Table~\ref{tabtemp} provides the derived rotational
temperatures of methanol lines and implied gas temperature.
A rotational diagram of the 165~GHz lines is presented in
Fig.~\ref{figdiagrot165}. All groups of lines (at 165~GHz, 213-220~GHz,
242~GHz, 252~GHz and 254-266~GHz) provide rotational temperatures $T_{rot}$
in very good agreement in the 18-24 K range. Due to the radiative decay of the
rotational levels within the ground vibrational state, we
expect colder rotational temperatures for the 242~GHz lines. A higher
collisional rate would limit this decay and observations suggest that indeed
the collision rate might be higher than modelled,
possibly due to the outgassing in narrow jets resulting in higher local
densities. Nevertheless we adopted a gas temperature $T_{gas}$ = 23 K
to derive all the molecular production rates.
We note also that if we use the "jet" part (velocity interval -0.8 to
-0.2~\kms) of the 165~GHz lines observed at high spectral (40~kHz) resolution
we find a temperature slightly higher by 3.2 K.
For comparison, at similar heliocentric distances, derived values are
$T_{gas}=30-40$~K for Hale-Bopp \citep{Biv02}, 18~K for C/2006~W3 (Christensen)
\citep{Boc10}, and 16~K for C/2002~T7 (LINEAR).

\begin{table}
\caption[]{Methanol rotational and derived kinetic temperatures.}\label{tabtemp}\vspace{-0.2cm}
\begin{center}
\begin{tabular}{lcccc}
\hline\hline\noalign{\smallskip}
UT date  & Freq. range  & lines   & $T_{rot}$ & $T_{gas}$ \\
$($mm/dd.dd) & (GHz)  & \tablefootmark{a}  & (K)  & (K) \\
\hline\noalign{\smallskip}
01/23.81 & 241-244\tablefootmark{b} & 10     & $17.9\pm2.4$ & $\ge25$  \\
01/24.01 &   165   &     7     & $19.2\pm3.9$ & $21.3\pm4.0$  \\
01/24.2  & 254-267 &     4     & $18.5\pm2.5$ & $18.6\pm2.6$  \\
01/24/2  & 251-252 & $6\times2$ & $21.8\pm7.1$ & $24.2\pm7.9 $  \\
01/24.78 & 213-230 &     6     & $20.7\pm3.0$ & $20.6\pm3.0$  \\
\hline
\end{tabular}
\end{center}
\tablefoot{
\tablefoottext{a}{Number of lines used for the determination of $T_{rot}$.
  Since the weighting is done according to the S/N of individual lines,
  if we take into account the
  lines below the formal detection threshold of $3-\sigma$,
  the derived $T_{rot}$ is not significantly changed.}
\tablefoottext{b}{These lines are more sensitive to the collision rate than to the gas
temperature itself: for a larger collision rate, the inferred $T_{gas}$ would
be lower.}
}
\end{table}

We can also constrain the kinetic temperature $T_{gas}$ from the CO(2-1)
line profile. Assuming that the width of the narrow component of the CO line
is due to thermal broadening, its full width at half maximum ($FWHM$)
of $0.237\pm0.003$~\kms~ implies
$T_{gas}\leq34\pm1$ K. This is compatible with $T_{gas}=23$ K derived from
the rotational temperatures. Modelling of the CO line shape (following section)
yields information on the actual dispersion of velocities. 

\subsubsection{Outgassing pattern and expansion velocity}
\begin{figure}[ht]
\centering
\resizebox{\hsize}{!}{\includegraphics[angle=270]{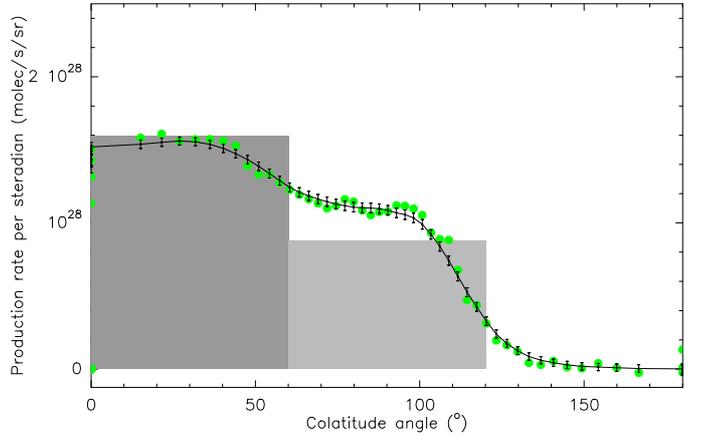}}
\caption{Inversion of the CO(2-1) line profile from Fig. \ref{figcocenter},
  assuming a constant expansion velocity of 0.56~\kms. The production rate
  per steradian is derived as a function of the colatitude angle ($\phi$=0\deg
  on the Earth side, 180\deg on the opposite), assuming symmetry along
  the Comet-Earth line. The dots correspond to the values derived for each
  spectral channel and the connected line with error bars is the running
  average taking into account thermal broadening.
  The grey shaded region corresponds to the production rate profile used to
  determine the total production rates,
  with $v_{exp}=0.56$~\kms~ and 0.50~\kms~ in the dark and light grey regions,
  respectively (see text).}
\label{figinverseco}
\end{figure}

        After determining the gas temperature and implied thermal broadening of
the lines, we modelled the CO line profile to infer the 
gas expansion velocity and outgassing pattern.
For $T_{gas}=20-23$ K the CO line shape asymmetry is well explained assuming
an expansion velocity $v_{exp}=0.56\pm0.01$~\kms~ on the Earth side and
considering that most of the outgassing is restricted to a cone with a
half-opening angle of 60\deg.
Figure~\ref{figinverseco} shows the CO production
rate per unit solid angle (assuming symmetry around the Comet-Earth axis)
as a function of the Earth-Comet-$V$gas or colatitude angle ($\phi$)
inferred from the line profile. For this profile inversion
(i.e. conversion of each channel intensity into a production rate
per solid angle, with the channel velocity $v_i$ converted into a
colatitude angle $\phi_i$ from the relationship  $v_i=-v_{exp}\cos(\phi_i)$),
we assumed that $v_{exp}$ (0.56~\kms) does not depend on $\phi$.
In reality, we expect the expansion velocity to be lower at 
higher phase angle due to lower surface temperature.
This is consistent with the smaller blue-shift (by 0.06~\kms) observed in
spectra acquired on the anti-sunward side (Fig.~\ref{figcooffset}).

\begin{figure}[ht]
\centering
\resizebox{\hsize}{!}{\includegraphics[angle=270]{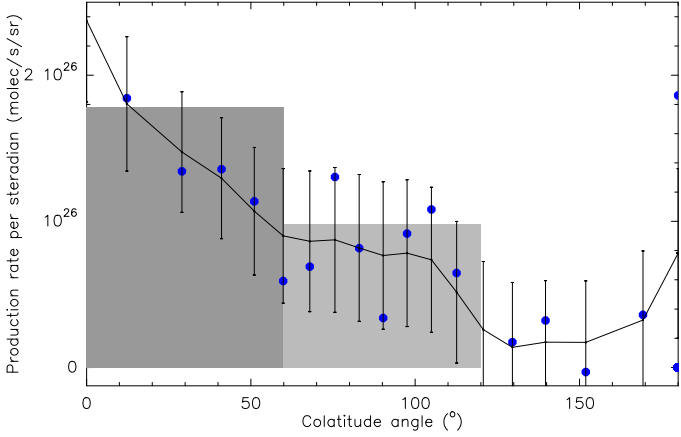}}
\caption{Inversion of the average of four CH$_3$OH lines (J=1 to 4) at 165~GHz.
  Scales are as in Fig.~\ref{figinverseco}.}
\label{figinversemet}
\end{figure}

In any case, line profile inversion with a lower velocity down to
0.3~\kms~ still produces a cut-off of production around the colatitude angle
$\phi$= 120-130\deg.
To approximately simulate the profile of gas production that was to be used
to derive total production rates, we assumed constant production rates
per solid angle $q_1$ and $q_2$ in the colatitude ranges $\phi_1=0-60$\deg
and $\phi_2=60-120$\deg, respectively. The expansion velocity was set to
0.56~\kms~ ($\phi_1=0-60$\deg) and 0.50~\kms ($\phi_2=60-120$\deg),
respectively.
We adjusted $q_1$ and $q_2$ to account for the observed mean Doppler shift
of the line. $q_2$ is about half $q_1$ and the total production rates in
these two regions ($Q_i = 2\pi\int_{\phi}q_i d\phi$) are in the ratio
$Q_1$:$Q_2$=10:11. The profile of the total production rate per solid angle
$q$ is plotted in Figs.~\ref{figinverseco} and \ref{figinversemet}.
The resulting line profiles are superimposed to the observed line profiles in
Figs.~\ref{figcocenter} and \ref{figcooffset}.

Although the agreement between simulation and observation is relatively good for
the central position (Fig.~\ref{figcocenter}), small differences appear at 
offset positions (Fig.~\ref{figcooffset}).
This shows that the actual outgassing pattern deviates somewhat from the
axisymmetric description done in this study.
Indeed the line intensity spatial distribution
(Figs.~\ref{figmapco} and \ref{figimgco}) points at some
asymmetry in the outgassing.

A similar profile inversion was undertaken for the average of the four strongest
methanol lines ($J$=1 to 4) observed at 165~GHz. The result
(Fig.~\ref{figinversemet}) is similar to the one from CO, although with
a lower S/N, suggesting that CH$_3$OH and CO have similar
production patterns from the nucleus. However the peak of the methanol
lines appears at slightly more negative velocities, possibly linked to the
uncertainty on the absolute line frequencies. Indeed the shift is $\sim50$kHz,
which is on the order of the differences between available frequencies in 2016
and previously published values \citep[][and references therein]{CDMS}.
If we expect the same line shape for CH$_3$OH and CO, the inferred CH$_3$OH line
frequencies are $165050.258\pm0.016$,
$165061.153\pm0.023$, $165099.220\pm0.017$ and $165190.544\pm0.010$ MHz,
for the $1_1-1_0$E, $2_1-2_0$E, $3_1-3_0$E, and $4_1-4_0$E lines,
respectively.

\subsubsection{Molecular production rates}

In summary, to analyse all spectra, we used a gas temperature of 23 K and the
outgassing pattern described above. A simpler description of the outgassing
pattern assuming hemispheric outgassing at $v_{exp}=0.56$~\kms~
yields production rates that are only 5\% higher. Production rates are 
provided in Tables~\ref{tabobsqp} and \ref{tabobsoff} and plotted in Fig. \ref{figqpdt}.
They are calculated as described in \citet{Biv16} and previous papers.
Results from the maps are provided in Table~\ref{tabobsoff} which
gives average intensities
at several radial offsets from the peak of intensity
determined from the maps (Figs.~\ref{figmapco} and \ref{figimgco}) for
the strongest lines and the derived production rates. There is only a marginal
trend of increasing production rate with offset for CO ($\sim+20$\%),
not significant for the other molecules, which could also be related to
instrumental effects (error beam and distortion of the beam shape at
higher elevations). For the three molecules, observations are compatible
with most of the production coming from the nucleus.

\subsection{Search for OH}
In order to constrain the water production rate of the comet, the 18-cm OH
lines were observed with the Nan\c{c}ay radio telescope.  From 2 January to
26 March, 2018, 48 individual daily spectra of about one-hour integration were
secured\footnote{Data are available in the public Nan\c{c}ay
database of cometary observations at
http://www.lesia.obspm.fr/planeto/cometes/basecom/}.
The observing and reduction procedures were as explained in
\citet{Cro02}.  During the
observing period, the OH maser inversion varied from $-0.31$ to $-0.15$.
Neither the individual spectra nor their average show any significant
signal.  For the 2 January -- 9 February average (corresponding to a mean
maser inversion of $-0.31$), the 3-sigma upper limit on the 1667 MHz line
area, integrated over a 3~\kms~ width, is 12 mJy km s$^{-1}$.
According to the model of \citet{Cro02}, this corresponds to an
upper limit of $1.1 \times 10^{28}$~\mols~ for the OH
production rate, that is, $1.2 \times 10^{28}$~\mols~ for the water
production rate.
Observations and resulting upper limits are summarised in
Table~\ref{tablognancay}.

\subsection{N$_2^+$/CO$^+$ from visible spectra}
Following the work presented by \citet{Coc18}, we estimate here
the N$_2^+$/CO$^+$ column density ratio from the visible spectra
presented in Sect 2.3.
We used line frequencies from \citet{Lof77}, \citet{Kep04}, and
\citet{Mag86}. Fluorescence efficiencies ($L/N$ or g-factors) of N$_2^+$
are from \citet{Lut93} and those for CO$^+$ are from \citet{Mag86}.
We use those pertaining to the heliocentric velocity range
-10 to +25~\kms~ (see sect. 6.3).

Synthetic fluorescence spectra were computed assuming a width of
$\sim0.4$~nm for each $v'-v''$ band and an instrumental resolution of 1~nm.
A N$_2^+$/CO$^+$ abundance ratio of 6--8\% was used to obtain the simulation
plotted in Figs.~\ref{spec-hboussier} and \ref{spec-ebertrand}.
Band areas were computed to estimate the relative intensities
(Table~\ref{tabvisspec}).
Due to non-perfect relative calibration across the visible band and also
variable uncertainties due to removal of atmospheric lines (mostly light
pollution lines of Hg-I and high pressure sodium (Na-I) broad emission above
540~nm, especially for the spectrum of 22.9 December, 2017, in
Fig.~\ref{imgspec-hboussier}), we took the
weighted average of the column densities derived from each detected line.
The band observed around 427~nm is the sum of the N$_2^+$(0-1) band at 427.4~nm
and the CO$^+(2-0)\Pi_{3/2}$ band at 427.2~nm \citep[band head,][]{Mag86}.
In order to evaluate the possible contribution of the N$_2^+$(0-1) band,
we computed the signal of the CO$^+(2-0)\Pi_{3/2}$ using the average of the
column density derived from the nearby CO$^+(2-0)\Pi_{1/2}$ band at 424.9~nm
and the column density derived from all CO$^+$ bands.
The 424.9~nm band should be less affected
by relative calibration issues, but the average of all lines should average
out local fluctuations of the baseline and uncertainties on the g-factors.

The derived N$_2^+$/CO$^+$ column density ratio is $0.09\pm0.03$ for 22.9 December
2017 and $0.06\pm0.02$ for 22.8 February 2018, giving an average of $7\pm2$\%.
If we only take into account lines below 460~nm to limit 
biases with wavelength, the values become
$0.08\pm0.02$ and $0.07\pm0.03,$ respectively, yielding an average of $8\pm2$\%.
Since ionization efficiencies are similar for the two molecules
($\beta_{i,N_2}=3.52\times10^{-7}$s$^{-1}$ and
$\beta_{i,CO}=3.80\times10^{-7}$s$^{-1}$ at 1 AU for quiet Sun \citep{Hue92}),
this translates to a N$_2$/CO production rate ratio of $8\pm2$\%,
consistent with the value of 6\% determined by \citet{Coc18}.

\begin{table}
  \renewcommand{\tabcolsep}{0.10cm}
\caption[]{N$_2^+$ and CO$^+$ bands from visible spectra.}\label{tabvisspec}\vspace{-0.5cm}
\begin{center}
\begin{tabular}{clccc}\hline\hline
$\lambda$   & band   & $L/N$\tablefootmark{a}  & \multicolumn{2}{c}{Col. Density\tablefootmark{b}} \\
      (nm)  &        & (10$^{-20}$W) & 22 Dec. & 22 Feb.  \\
\hline
 398--403 & CO$^+(3-0)$    & 0.215      & $1000\pm102$& $1000\pm45$ \\
 423--426 & CO$^+(2-0)^B$\tablefootmark{c}
                           & 0.100      & $901\pm173$ & $1721\pm47$ \\
 451--458 & CO$^+(4-2)$    & 0.145     &  $879\pm44$ & $910\pm33$ \\
                & +CO$^+(1-0)$  & & & \\
 468--474 & CO$^+(2-1)$    & 0.085     &  $809\pm61$ & $1636\pm56$ \\
 482--488 & CO$^+(3-2)$    & 0.044     &  $296\pm117$ & $1125\pm106$ \\
 502--510 & CO$^+(1-1)$    & 0.123     &  $610\pm51$ & $1610\pm38$ \\
\hline
 \multicolumn{3}{l}{Average CO$^+$ column density:}& $768\pm184$& $1320\pm377$\\
\hline
 388--393 & N$_2^+(0-0)$   & 3.56      &   $73\pm7$ &  $87\pm16$  \\
 426--429 & CO$^+(2-0)^R$\tablefootmark{c} & 0.115 & {\it $835\pm179$}\tablefootmark{d} & {\it $1520\pm200$}\tablefootmark{d} \\
          & +N$_2^+(1-2)$    & 1.02    &   $77\pm29$\tablefootmark{e} & $50\pm38$\tablefootmark{e} \\
\hline
\multicolumn{3}{l}{Average N$_2^+$ column density:}&   $73\pm7$  & $83\pm14$ \\
\hline
\hline
\multicolumn{3}{l}{Col. density ratio N$_2^+$/CO$^+$~\tablefootmark{f}:} & $0.09\pm0.03$ & $0.06\pm0.02$ \\
\hline
\end{tabular}
\end{center}
\tablefoot{
\tablefoottext{a}{Values at 1 AU for N$_2^+$ \citep{Lut93}, and for
  CO$^+$ \citep{Mag86} for (-10, +30~\kms) heliocentric velocity interval.}
\tablefoottext{b}{in relative units, normalized to the first CO$^+$ band:
integrated band intensity (in $ADU\times\AA$) on the spectrum divided by $L/N$.}
\tablefoottext{c}{$^B$: CO$^+(2-0)\Pi_{1/2}$ blue wing and $^R$:
  CO$^+(2-0)\Pi_{3/2}$ red wing.}
\tablefoottext{d}{The intensity of the CO$^+(2-0)\Pi_{3/2}$ band is derived from
  this range of column density estimated from the nearby $\Pi_{1/2}$ band and
  the average of all bands.}
\tablefoottext{e}{The intensity of the N$_2^+(1-2)$ band is inferred from
  the subtraction of the expected contribution of the CO$^+(2-0)\Pi_{3/2}$.}
\tablefoottext{f}{Average column density ratio, including a 20\% additional
  uncertainty to each column density due to the use of approximate values of
  the $L/N$ parameter which varies with the heliocentric velocity of the ions \citep{Mag86}.}
}
\end{table}

\begin{table*}
\caption[]{Line intensities from IRAM observations and production rates (Upper limits are $3-\sigma$).}\label{tabobsqp}\vspace{-0.5cm}
  \renewcommand{\arraystretch}{0.87}
\begin{center}
\begin{tabular}{lllccl}
\hline\hline
Date & Molecule & Transition         & Frequency & Intensity & Total production rate \\
$[yyyy/mm/dd.dd]$ &  &               & [GHz]     & [mK~\kms] &[$10^{26}$~molec.s$^{-1}$] \\
\hline
2018/01/23.79 & CO       & 2-1          & 230.538 & $1109\pm15$ & \hspace{0.2cm}$951\pm13$ \\
2018/01/23.80 & CO       & 2-1          & 230.538 & $1108\pm25$ & \hspace{0.2cm}$969\pm22$ \\
2018/01/23.82 & CO       & 2-1          & 230.538 & $1077\pm24$ & \hspace{0.2cm}$929\pm21$ \\
2018/01/23.83 & CO       & 2-1          & 230.538 & $1109\pm18$ & \hspace{0.2cm}$957\pm16$ \\
2018/01/24.72 & CO       & 2-1          & 230.538 & $1137\pm18$ & $1131\pm18$ \\
2018/01/24.73 & CO       & 2-1          & 230.538 & $1061\pm16$ & $1056\pm16$ \\
2018/01/24.76 & CO       & 2-1          & 230.538 & $1346\pm22$ & $1098\pm18$ \\
2018/01/24.77 & CO       & 2-1          & 230.538 & $1415\pm19$ & $1141\pm15$ \\
2018/01/24.78 & CO       & 2-1          & 230.538 & $1391\pm19$ & $1122\pm15$ \\
2018/01/24.82 & CO       & 2-1          & 230.538 & $1172\pm17$ & \hspace{0.2cm}$956\pm14$ \\
2018/01/24.83 & CO       & 2-1          & 230.538 & $1239\pm16$ & \hspace{0.2cm}$989\pm13$ \\
2018/01/24.85 & CO       & 2-1          & 230.538 & $1138\pm19$ & \hspace{0.2cm}$957\pm16$ \\
\hline
2018/01/24.78 & $^{13}$CO & 2-1           & 220.399 &    $10\pm8$ & \hspace{0.1cm} $<21$ \\
\hline
2018/01/24.80 & CO+ & 2-1 $F=3/2-1/2$     & 235.790 &    $203\pm51$\tablefootmark{a} & \\
2018/01/24.80 & CO+ & 2-1 $F=5/2-3/2$     & 236.063 &    $227\pm46$\tablefootmark{a} & \\
\hline
2018/01/23.81 & CH$_3$OH & $5_{0}-4_{0}$A   & 241.791 & $123\pm13$ & \vline \\
              &          & $5_{-1}-4_{-1}$E & 241.767  & $97\pm13$ & \vline \\
              &          & $5_{0}-4_{0}$E   & 241.700  & $79\pm13$ & \vline \\
              &          & $5_{+1}-4_{+1}$E & 241.830  & $31\pm12$ & \vline\hspace{0.1cm} $11.1\pm2.1$\tablefootmark{b}\\
              &          & $5_{2}-4_{2}$E   & 241.904  & $53\pm13$ & \vline \\
              &          & $5_{1}-4_{1}$A$^-$ & 243.916 & $52\pm15$ & \vline \\
              &          & $5_{2}-4_{2}$A$^-$ & 241.844 & $ 9\pm12$ & \vline \\ \vspace{0.1cm} 
              &          & $5_{2}-4_{2}$A$^+$ & 241.888 & $14\pm14$ & \vline \\ 
2018/01/24.01 & CH$_3$OH & $1_{+1}-1_{0}$E & 165.050 & $24\pm 7$ & \vline \\
              &          & $2_{+1}-2_{0}$E & 165.061 & $35\pm 7$ & \vline \\
              &          & $3_{+1}-3_{0}$E & 165.099 & $34\pm 7$ & \vline \\
              &          & $4_{+1}-4_{0}$E & 165.190 & $28\pm 7$ & \vline\hspace{0.1cm} $10.2\pm1.1$\\
              &          & $5_{+1}-5_{0}$E & 165.369 & $12\pm 8$ & \vline \\
              &          & $6_{+1}-6_{0}$E & 165.679 & $14\pm 9$ & \vline \\
              &          & $7_{+1}-7_{0}$E & 166.169 & $ 3\pm 8$ & \vline \\

2018/01/24.01 & CH$_3$OH & $3_{+2}-2_{+1}$E & 170.061 & $51\pm9$ & \hspace{0.1cm} $12.0\pm2.1$ \\

2018/01/24.2  & CH$_3$OH & $J_{3}-J_{2}$A$^{\pm}$  & 251.5-252.0 & $117\pm28$\tablefootmark{c} & \vline \\
2018/01/24.2  & CH$_3$OH & $2_{0}-1_{-1}$E  & 254.015 & $45\pm 8$ & \vline \\
2018/01/24.2  & CH$_3$OH & $5_{+2}-4_{+1}$E & 266.838 & $46\pm10$ & \vline\hspace{0.1cm} $14.9\pm2.6$ \\
2018/01/24.2  & CH$_3$OH & $6_{+1}-5_{+2}$E & 265.290 & $23\pm10$ & \vline \\
2018/01/24.2  & CH$_3$OH & $9_{0}-8_{+1}$E  & 267.403 & $ 1\pm10$ & \vline \\

2018/01/24.5  & CH$_3$OH & $8_{-1}-7_{0}$E  & 229.759 & $ 5\pm 5$ & $<15.8$ \\
2018/01/24.5  & CH$_3$OH & $3_{-2}-4_{-1}$E & 230.027 & $11\pm 5$ & $<22.4$ \\

2018/01/24.78 & CH$_3$OH & $1_{+1}-0_{0}$E & 213.427 & $28\pm 8$ & \vline \\
2018/01/24.78 & CH$_3$OH & $5_{+1}-4_{2}$E & 216.946 & $10\pm 8$ & \vline\hspace{0.1cm} $11.8\pm1.5$\\
2018/01/24.78 & CH$_3$OH & $4_{+2}-3_{1}$E & 218.440 & $52\pm 6$ & \vline \\
2018/01/24.78 & CH$_3$OH & $8_{0}-7_{+1}$E & 220.078 & $ 1\pm 7$ & \vline \\

\hline
2018/01/23.81 & H$_2$CO & $3_{1,2}-2_{1,1}$ & 225.698 & $52\pm11$ & $0.54\pm0.10$\tablefootmark{b} \\
2018/01/24.01 & H$_2$CO & $2_{1,1}-1_{1,0}$ & 150.498 & $13\pm 7$ & $0.30\pm0.16$ \\
2018/01/24.78 & H$_2$CO & $3_{0,3}-2_{0,2}$ & 218.222 & $24\pm 6$ & $0.43\pm0.10$\tablefootmark{b} \\
\hline
2018/01/24.3  & HCN      & $3-2$       & 265.886  & $28\pm10$ & $0.04\pm0.01$ \\
\hline
2018/01/24.01 & H$_2$S & $1_{1,0}-1_{0,1}$ & 168.762 & $<25$ & $<0.71$ \\
\hline
2018/01/23.81 & CS & $5-4$ & 244.936 & $<38$ & \vline \\
2018/01/24.01 & CS & $3-2$ & 146.969 & $<20$ & \vline\hspace{0.1cm}  $<0.14$ \\
\hline
2018/01/24.01 & CH$_3$CN & $(8,0-7,0)+(8,1-7,1)$ & 147.173 & $14\pm 7$ & \vline \\
2018/01/24.01 & CH$_3$CN & $(9,0-8,0)+(9,1-8,1)$ & 165.568 & $10\pm11$ & \vline\hspace{0.1cm} $<0.12$ \\
\hline
2018/01/24.01 & HC$_3$N & $17-16 + 18-17$ & 154.66 \& 163.75 & $<31$ & $<0.56$ \\
\hline
2018/01/24.01 & HNCO & $7_{0,7}-6_{0,6}$ & 153.865 & $17\pm 7$ & \vline \\
2018/01/24.78 & HNCO & $10_{0,10}-9_{0,9}$ & 219.798 & $19\pm 8$ & \vline\hspace{0.1cm}  $0.50\pm0.16$ \\
2018/01/23.81 & HNCO & $11_{0,11}-10_{0,10}$ & 241.774 & $ 2\pm10$ & \vline \\
\hline
2018/01/24.2  & NH$_2$CHO & 9 lines & 162.96--267.06 &  -  & $<0.34$ \\
\hline
2018/01/24.2  & CH$_3$CHO & 16 lines & 149.50--244.83 &  -  & $<0.99$ \\
\hline
2018/01/24.2  & HCOOH & 7 lines & 151.18--252.08 &  -  & $<1.83$ \\
\hline
2018/01/24.78 & SO   & $(5,5)-(4,4)$ & 215.220 & $ 2\pm 6$  & \vline \\
2018/01/24.78 & SO   & $(5,6)-(4,5)$ & 219.949 & $-7\pm 8$  & \vline\hspace{0.1cm}  $<1.5$\tablefootmark{d} \\
2018/01/24.2  & SO   & $(6,5)-(5,4)$ & 251.826 & $ 1\pm 8$  & \vline \\
\hline
2018/01/24.2  & SO$_2$ & 10 lines & 162.96--267.06 &  -  & $<0.34$ \\
\hline
2018/01/24.3  & PH$_3$      & $1-0$     & 266.944  & $<31$ & $<0.14$ \\
\hline
\end{tabular}
\end{center}
\tablefoot{
\tablefoottext{a}{The line integration window is -5 to +30~\kms.}
\tablefoottext{b}{Average production rate for the period also taking into
  account measurements at 2 to 10\arcsec~ offsets (details in
  Table~\ref{tabobsoff}). For formaldehyde we assumed that all molecules come
  from the nucleus.}
\tablefoottext{c}{Sum of 12 lines (J=3 to 8).}
\tablefoottext{d}{SO is assumed to come from the photo-dissociation of SO$_2$
  with a scale-length of 17000~km (at $r_h=2.8$ AU).}
}\\
\end{table*}

\begin{table*}
\caption[]{Line intensities and production rates based on mapping data.}\label{tabobsoff}
\begin{center}
\begin{tabular}{lllcccc}
\hline\hline
Date & Molecule & Transition         & offset\tablefootmark{a} & Intensity\tablefootmark{a} & Total production rate \\
$[yyyy/mm/dd.dd]$ &  &               & [\arcsec]     & [mK~\kms] &[$10^{26}$~molec.s$^{-1}$] \\
\hline
2018/01/23.81 & CO       & 2-1          &  3.0 & $1132\pm12$ &   $977\pm10$ \\
              &          &              &  7.9 & $ 776\pm32$ &  $1100\pm45$ \\
              &          &              & 11.9 & $ 537\pm44$ &  $1219\pm100$ \\
2018/01/24.80 & CO       & 2-1          &  1.7 & $1285\pm 8$ &  $1040\pm 6$ \\
              &          &              &  4.7 & $1110\pm13$ &  $1104\pm13$ \\
              &          &              &  8.5 & $ 882\pm34$ &  $1351\pm52$ \\
              &          &              & 12.3 & $ 501\pm26$ &  $1192\pm62$ \\
              &          &              & 15.4 & $ 405\pm47$ &  $1288\pm150$ \\
\hline
2018/01/23.81 & CH$_3$OH & $5_{K}-4_{K}$ (6 lines) & 3.0 & $435\pm32$\tablefootmark{b} & $11.1\pm2.1$ \\
              &          &              &  7.9 & $233\pm83$\tablefootmark{b} & $11.4\pm4.1$ \\
\hline
2018/01/23.81 & H$_2$CO & $3_{1,2}-2_{1,1}$ & 3.0   & $52\pm11$ & $0.53\pm0.11$ \\
              &         &                &  7.9  & $39\pm28$ & $0.88\pm0.63$ \\
2018/01/24.78 & H$_2$CO & $3_{0,3}-2_{0,2}$ &  2.9 & $24\pm 6$ & $0.42\pm0.10$ \\
              &         &                &  8.6  & $13\pm21$ & $0.58\pm0.94$ \\
\hline
\end{tabular}
\end{center}
\tablefoot{
  \tablefoottext{a}{Mean radial offset from the position of peak intensity and
    corresponding line integrated area, for radial averages of mapping points.}
  \tablefoottext{b}{Sum of the six strongest lines.}
}
\end{table*}

\subsection{Dust production}

Our analysis on the dust production of the comet is solely based on optical
data, as the detection of continuum emission in the infrared or in the radio
has not been reported so far. The optical spectra present only a weak
continuum (Fig.~\ref{imgspec-hboussier}). Images show a ``CO$^+$'' blue coma,
and do not reveal any dust tail \citep{Coc18,dbcomet}, suggesting a low dust
production. From coma photometry, F. Kugel and H. Boussier \citep{dbcomet}
report $Af\rho$ values from 530 to 830~cm, with a mean value of 670~cm for the period from
8 February to 8 March, 2018 (at a nearly constant phase angle of 21\deg).
The $Af\rho$ parameter introduced by \citet{Ahe84} is proportional to the
dust loss rate multiplied by the cross-section of the dust particles within the
aperture; it can be used as a proxy for the dust production. 
According to \citet{Ahe95}, a value of 1000~cm
corresponds to a dust production rate of about 1 ton per second. For
comet C/2016~R2 this would suggest that the dust production is
about one order of magnitude less than the total gaseous production, in mass
($Af\rho/Q_{\rm gas} \approx 0.1$ cm.kg$^{-1}$ vs. 2--5 for other comets
at 3 AU (Table~\ref{tabqp3au}) and $Af\rho/Q_{\rm H_2O} = 0.5$ cm.kg$^{-1}$
on average from \citet{Ahe95} for comets closer to the sun).
The conversion of the $Af\rho$ parameter into dust production rate is highly
dependent on assumptions made on the dust size distribution and maximum size.
If released dust consists predominantly of large grains, then the dust
production rate could be much higher than the estimation given above. These
large grains could form a dust trail which is not seen in the images,
including those acquired in mid-December 2017 when the Earth crossed the plane
of the orbit of the comet. As discussed in the following section, 
the $Af\rho$ value for C/2016 R2 is comparatively much lower than values
measured in comets showing comparable gaseous activity.


\section{Discussion}

The derived CO production rate ($10.6\times10^{28}$~\mols, equivalent
to $\sim5$ tons per second) corresponds to a large outgassing rate,
only surpassed by comet Hale-Bopp and by 17P/Holmes during its massive outburst
(Table~\ref{tabqp3au}). This was
unexpected given the small brightness of the comet (total visual magnitude
m$_1$ around 10--11). Indeed, from the correlation between visual magnitudes
and CO production \citep{Biv01} we obtain a CO production rate
20 times lower. This indicates a dust-poor coma, consistent with the 
very low $Af\rho$ values ($670\pm110$~cm) measured by F. Kugel and H. Boussier
during this period. At the same heliocentric distance for comet Hale-Bopp
the $Af\rho$ was 100 times higher \citep{Wei03} (at a similar phase angle)
for a CO production only twice higher.
We note that the visual magnitude is difficult to compare to that of
other comets as the optical brightness is dominated by CO$^+$ "tail" emission
lines.

\subsection{Molecular abundances}
\begin{figure}
\centering
\resizebox{\hsize}{!}{\includegraphics[angle=0]{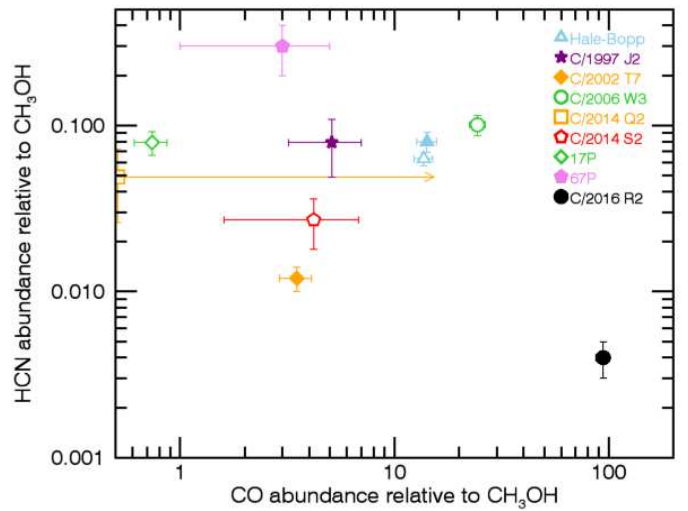}}
\caption{CO/CH$_3$OH and HCN/CH$_3$OH ratios in comets observed between
  2.3 and 3.3 AU from the Sun.}
\label{fig3aucohcn}
\end{figure}

Pending the assessment of the abundance of other possible major molecules
(CO$_2$, O$_2$), for which no ionic emission has been reported so far,
the major species in the coma of comet C/2016~R2 (PanSTARRS) at 2.8 AU from
the Sun are CO and N$_2$ with an abundance ratio on the order of $100:8$.
Observations and searches for NH$_3$ or NH$_2$ and hydrocarbons are needed
to complete the inventory, but these species are not expected to be as abundant
as the
previous species. Water is not detected with an upper limit H$_2$O/CO $<$ 0.1.
Comparing with comets observed at similar heliocentric distances
(Table~\ref{tabqp3au}), methanol is relatively abundant relative to water,
but deficient compared to CO. Hydrogen cyanide  and sulfur species are strongly depleted.
Figure~\ref{fig3aucohcn} compares the abundances of
CO and HCN relative to methanol in comets observed at 2.3-3.3 AU from the Sun.
Molecular production rates measured in these comets are listed in
Table~\ref{tabqp3au}.

\subsection{Upper limits on minor species}
The wide frequency coverage (Table~\ref{tablog}) has enabled us to look for
many other species previously detected in comets \citep{Biv15}. The most
significant upper limits are provided in Table~\ref{tabobsqp}. For species
where several lines of similar expected intensities are sampled, we
provide the combined upper limit (weighted rms) derived from all lines.
In general the upper limits in comparison to the reference molecule
(HCN for CN-species,  CH$_3$OH for CHO-species and H$_2$S for
S-species) are not very constraining \citep{Biv15}.
The marginal detection of HNCO in C/2016~R2 requires confirmation as it
corresponds to a HNCO/CH$_3$OH ratio that is between three and ten times higher than in other comets
and a HNCO/HCN two orders of magnitude higher.
H$_2$S is clearly under abundant ($<6$\% relative to CH$_3$OH) in comparison
to other comets observed at $\sim$3 AU from the Sun (Table~\ref{tabqp3au}:
H$_2$S/CH$_3$OH = 30-70\%).

We also looked for the CO isotopologues $^{13}$CO and C$^{17}$O, but they were
not detected. The best S/N is obtained considering the blue-shifted part of
the lines, and gives $^{12}$C/$^{13}$C$>54$ ($82\pm42$ at 2~$\sigma$), which is
compatible with the terrestrial value (90).

\subsection{Ions observed with IRAM-30m}
     The rotational lines of CO$^+$ were observed for the first time in comet
Hale-Bopp at the CSO \citep{Lis97} and the IRAM 30-m \citep{Hen01}.
The CO$^+$ lines at 236~GHz are marginally detected,
red-shifted and broader than the CO line, as would be expected for ions
(cf Hale-Bopp spectra of CO$^+$ and HCO$^+$ \citep{Lis97}).
The two strongest CO$^+(N=2-1)$ transitions at 236062.553 and 235789.641~MHz
show a marginal $3-2~\sigma$ peak of FWHM=1~\kms~ close to the zero velocity
in the comet frame and a broader component of  25~\kms~ in width between -5 and +30~\kms~ more clearly detected (5 and 4~$\sigma$,
respectively). This asymmetry towards larger red-shifted velocities is due
to the acceleration of ions by the solar wind in the anti-solar direction.
Figure~\ref{figcoplus} shows the combined spectrum (average of the two lines).
Column densities cannot be derived in a straightforward manner as the
properties of this profile show evidence of the acceleration of CO$^+$ ions
in the beam due to interaction with the solar wind. A rough estimate
for an expansion velocity of 10~\kms~ (mean Doppler shift of the lines),
and a rotational population at 23 to 200~K
is 0.4--1.7$\times10^{12}$cm$^{-2}$, but we have no precise idea of the
rotational population of CO$^+$ ions which are in a collisionless regime.
We assumed that CO$^+$ comes from the photo-ionization of CO (scale-length on
the order of $12\times10^6$~km) which yields optically thin lines in any case.
The marginal narrow component would yield a column density about five times
lower.
However, this profile can be used to obtain more appropriate g-factors to
analyse the optical spectra taken in a similar aperture: given the heliocentric
velocity of the comet and the small phase angle of the observations, this means
that the heliocentric velocity of the CO$^+$ ions contributing the most of the
signal is in the -10 to +25~\kms~ range.
Hence we used the corresponding $L/N$ values from \citet{Mag86}
to interpret the optical spectra.

HCO$^+$ is not detected. It has been detected in comets with weaker
productions of CO, but at closer distances to the Sun where water production
and protonation of CO in the coma are much more efficient \citep{Mil04}.

\subsection{Variation with time}
\begin{figure}
\centering
\resizebox{\hsize}{!}{\includegraphics[angle=0]{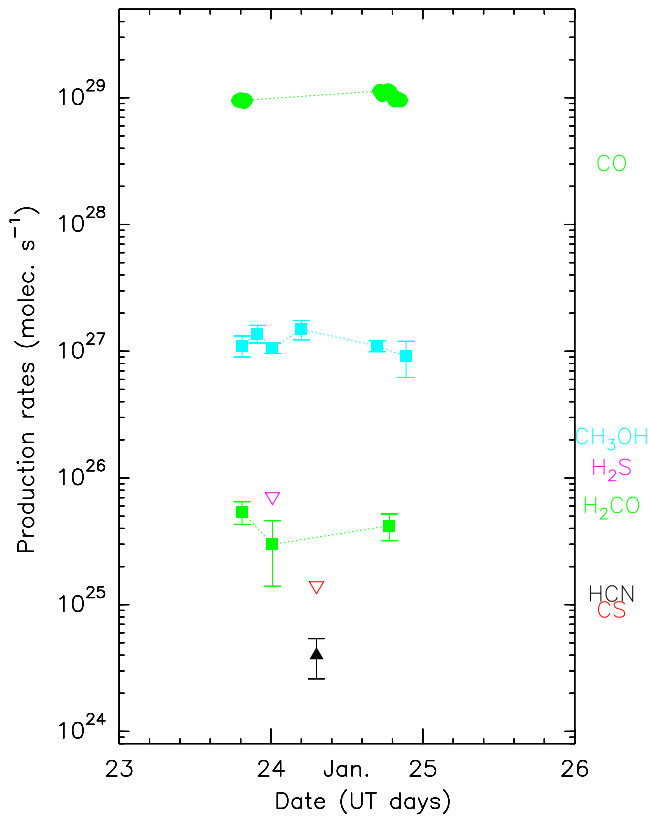}}
\caption{Evolution of the production rates (and upper limits) measured in comet
C/2016~R2 (PanSTARRS) during the time interval spanning 23 and 24 January, 2018.}
\label{figqpdt}
\end{figure}

 Since the comet exhibited a unexpectedly large outgassing of CO, with
some optical images showing rapid changes in the CO$^+$ structures, we looked
for possible evidence of short-term variations due to either the rotation of the
nucleus or a transient outburst phase.
   Figure~\ref{figqpdt} shows the production rates from Table~\ref{tabobsqp} and
their evolution over the two days of observations. On this short time scale,
generally the variations were less than 20\% and possibly more related to
pointing and calibration uncertainties. Indeed, at high elevations
(above 60--70\deg) the beam efficiency of IRAM-30m degrades. We made some
modelled corrections of the beam efficiency but could not precisely track its
variation. This may account for $\sim$10\% variations.
No significant variations of the CO(2-1)
line Doppler shift ($-0.28\pm0.02$~\kms) that could be correlated with
a variation in the production pattern -- and independently of any
calibration issue -- are observed either.
So, on a  timescale of several hours, we do not see any significant
variation of the activity.

\begin{table*}
  \renewcommand{\tabcolsep}{0.13cm}
\caption[]{Production rates of comets around 2.8 AU from the Sun.}\label{tabqp3au}
\begin{center}
\begin{tabular}{llccccccccc}
\hline\hline
Comet & $r_h$\tablefootmark{a} & $Q_{\rm CH_3OH}$ & $Q_{\rm HCN}$ & $Q_{\rm CO}$ & $Q_{\rm H_2CO}$ & $Q_{\rm H_2S}$ & $Q_{\rm CS}$ & $Q_{\rm H_2O}$ & $Af\rho$ & Ref.\\
      & [AU] & \multicolumn{7}{c}{[$\times10^{26}$\mols]} & [cm] & \\
\hline
Hale-Bopp & -2.84 &  $141\pm 5$ & $11.5\pm1.5$ & $2000\pm200$ & $5.5\pm0.5$ & $100\pm10$ & $6\pm2$ & $6000\pm500$ & $\sim80000$ & 1 \\
          & 2.88 &  $100\pm10$ & $ 6.3\pm0.1$ & $1370\pm 40$ & $4.5\pm0.5$ & $ 61\pm 4$ & $3\pm1$ & $1500\pm700$ & $\sim50000$ & 1 \\
C/1997~J2 &  3.05 & $7.6\pm2.1$ & $ 0.6\pm0.2$ & $39\pm10$   &              &            &          &              &  & \\
C/2002~T7 & -2.74 &  $25\pm1$   & $0.29\pm0.04$ & $87\pm15$  & $<3.0$       &            & $<4.8$ & $280\pm20$ & $\sim2500$ & 6 \\
17P/Holmes  & 2.44 & $2150\pm250$ & $170\pm20$ &  $1600\pm200$ &             &         & $128\pm2$ & $\sim33000$ & $\sim500000$ & 2 \\
          & 2.45 &  $350\pm100$ &  $21\pm2$  &               &             &            & $10.7\pm0.1$ & $\sim8700$ & \\
          & 2.46 &  $250\pm30$  &  $12\pm2$  &               & $7.8\pm1.7$ & $ 80\pm 5$ & $ 8 \pm 1$ &            & \\
C/2006~W3  & 3.20 & $15.8\pm0.9$ & $1.6\pm0.2$ & $386\pm25$  & $<2.4$       & $10.1\pm1.0$ & $0.45\pm.19$ & $<200$ & $\sim3500$ & 3,4,6\\
C/2014~Q2 &  3.36 & $9.2\pm2.6$ & $0.45\pm0.17$ &   $<141$   &  $<1.5$     &            &  $<0.66$    &              & $\sim350$ & 6 \\
C/2014~S2 &  2.31 &  $15\pm4$   &  $0.4\pm0.1$  &  $63\pm36$ & $0.22\pm0.11$ &            &  $<0.29$    &              & $\sim1800$ & 5 \\
67P/C.G.  & -2.80 & $0.012\pm.006$ &             & $<0.018$   &             &             &            & $0.8\pm0.2$   & $4$ & 6 \\
C/2016~R2 &  2.83 & $11.2\pm0.7$ & $0.04\pm0.01$ & $1056\pm47$ & $0.45\pm0.07$ &$<0.71$   &  $<0.14$    &  $<120$  & $670$ & 5 \\
\hline
\end{tabular}
\end{center}
\tablefoot{
Unpublished production rates are based on ground-based observations
with IRAM-30m, CSO, JCMT or Nan\c{c}ay radio telescopes.\\
\tablefoottext{a}{Negative values mean pre-perihelion observations.}
}
\tablebib{
  (1)~\citet{Wei03}; (2) \citet{Sch09}; (3) $Q_{dust}=900$~kg/s, \citet{Boc10};
  (4) $Af\rho\approx10000$~cm and $Q_{dust}=380$ kg/s pre-perihelion at 3.1 AU, \citet{Kor16};
  (5) \citet{dbcomet}; (6) M. Kidger / Spanish observers comet home page (http://www.observadores-cometas.com/cometas.html);
}\\
\end{table*}

\section{Summary and conclusion}

We performed a multi-wavelength (optical, millimetre, centimetre) compositional
study of comet C/2016~R2 (PanSTARRS). This comet has the following characteristics.
\begin{itemize}
\item A very large CO production, on the order of 10$^{29}$~\mols~ (i.e.
  5 tons/s) at $r_h$=2.8 AU from the Sun, only a factor of two below that of
  C/1995 O1 (Hale-Bopp) at same $r_h$.
\item Very low dust production, based on the Af$\rho$ which is lower than
  values measured in comets of similar gaseous activity at $\sim$ 3 AU from
  Sun by more than a factor of 15.
\item Unusual optical spectrum showing essentially CO$^+$ and N$_2^+$ lines.
\item Abundant N$_2$ in the coma (N$_2$/CO$\sim$0.08), with N$_2$ dominating
  the nitrogen budget.
\item A strong depletion of H$_2$O, CH$_3$OH, H$_2$CO, HCN, H$_2$S relative to
  CO (by more than one order of magnitude) compared with other comets observed
  at a similar heliocentric distance.
\item A depletion of HCN and sulfur species relative to methanol.
\end{itemize}

The origin of the huge production of CO of C/2016~R2 could be related to
the size
of its nucleus. The nucleus of comet Hale-Bopp has a radius estimated to
$R_n$ $\sim$ 37~km \citep{Alt99}, making this comet one of the largest ever
observed. Distant comet 29P/Schwassmann-Wachmann 1, which also shows a large
CO production on its circular 6-AU orbit ($\sim4\times10^{28}$~\mols), has a
radius of $\sim$23~km \citep{Sta04}. Unfortunately, measurements of the nucleus
size of C/2016~R2 have not yet been reported. Given
the pure CO ice sublimation rate at 2.8 AU $Z_{\rm CO}(2.8)\sim1.6\times10^{22}$
molec.m$^{-2}$.s$^{-1}$ \citep{Sek91}, the outgassing rate of C/2016~R2 could
be supplied by a pure CO ice object of 3~km in diameter.
However, detailed thermal and structural modelling is necessary to provide
valuable constraints on the nucleus size from the CO production rate. 

The depletion of H$_2$O, CH$_3$OH, H$_2$CO, HCN, H$_2$S relative to CO,
together with the low dust production, suggest that the large CO activity
reflects a CO-rich ice composition and large ice/dust ratio within the 
nucleus material of C/2016~R2. Indeed, due to low latent heat, sublimation of CO ice can
release smaller particles and larger aggregates than the outgassing of H$_2$O
ice \citep{Gun15}. Therefore,  if C/2016~R2 had a dust/ice ratio comparable to
that of other comets, significant dust production should have been observed. 
Here, we assume that the thermal properties of the  nucleus surface and
subsurface layers of C/2016~R2, controlled by the heat conductivity and porosity of the
material, are overall comparable to other comets. This seems a reasonable
assumption. Indeed, the illumination-driven CO outgassing indicated by the CO
line profile (Sect. 3.1.2) shows that the CO sublimation front is close to the
surface, consistent with a low-porosity material. Other comets observed at
$r_h > $ 3 AU from the Sun (Hale-Bopp and 29P) display a similar CO outgassing
pattern \citep[e.g.][]{Gun03}.
In comet Hale-Bopp, which had a higher $Af\rho$/gas ratio and higher
abundance of H$_2$O and HCN at 2.8 AU (Table~\ref{tabqp3au}), icy grains
were also found in the coma \citep{Lel98}. We can speculate
that following the low dust abundance, icy grains are also deficient in the
coma of this comet, which could explain the low abundance of water and HCN,
which could be significantly produced from the sublimation of icy grains
in other comets like Hale-Bopp at such heliocentric distances.

The other particularity of comet C/2016~R2 is the large abundance of N$_2$,
dominating other N-bearing species. Measured N$_2$/HCN is $< 0.006$ for
Hale-Bopp \citep[derived from][using HCN/CO=0.01]{Coc18}, and $\sim$0.2 in 67P
at $r_h$ =3 AU  \citep[from][]{Rub15,Ler15}. For C/2016~R2, N$_2$/HCN = 2000,
that is, four orders of magnitude higher. Since there is no hint of NH$_2$
lines in optical
spectra, and other N-bearing species (Table~\ref{tabobsqp}) searched for at
IRAM are less abundant than N$_2$ by at least two orders of magnitude, we can
conclude that contrary to other comets, most of the nitrogen escaping the
nucleus is in the form of N$_2$.

Comet C/2016~R2 shows a remarkably high N$_2$/CO ratio ($\sim0.08$), comparable
to the largest values measured in comets so far. A detailed comparison with
other comets is given in
\cite{Coc18}.  Values are, for example, N$_2$/CO $<$ 6 $\times$ 10$^{-5}$ for comet
Hale-Bopp \citep{Coc00}, 5.7 $\times$ 10$^{-3}$ for 67P \citep{Rub15},
0.01 for 29P \citep[from a marginal detection of N$_2^+$,][]{Kor08},
and 0.06 in C/2002~VQ$_{94}$ (LINEAR) \citep{Kor14}.    

The strong nitrogen deficiency in some comets, both in the ice and refractory
phases, was revealed during the space investigation of 1P/Halley, and
confirmed with the Rosetta mission \citep{Rub15,Fra17}. Its interpretation
remains elusive. According to \citet{Owe95}, the trapping of N$_2$ by amorphous
water ice in the cooling solar nebula was inefficient within Neptune's orbit,
resulting in the formation of  planetesimals deficient in N$_2$. Laboratory 
experiments show that the trapped N$_2$/CO ratio is depleted by a factor
of $\sim$ 20 at 24 K, with respect to the gas phase value, the depletion factor
being strongly dependent on temperature \citep{Bar07}. With a protosolar
ratio N/C = 0.29 and assuming that all C and N is in the form of CO and N$_2$,
the N$_2$/CO ratio in the solar nebula gas phase is 0.15, resulting in a trapped
N$_2$/CO of a few times 10$^{-3}$ at $\sim$ 25 K, relatively consistent with the value
measured for 67P  \citep{Rub15}. If this interpretation is correct, the very
low N$_2$/CO values measured in, for example, comet Hale-Bopp would indicate a
formation in warmer regions of the solar nebula. 

Considering the gas-phase species detected in comet C/2016~R2, the N/C ratio
for this comet is $\sim$ 0.15, that is, close to the solar value of 0.29$\pm$0.10
\citep{Lod09}. To explain this property, a possibility is that C/2016~R2
agglomerated from grains formed at an extremely low temperature, favouring the
trapping of high quantities of N$_2$ both as trapped gas and in condensed form.
However, the high CO content of C/2016~R2 ices together with the low dust/ice
ratio (assuming our extrapolation from coma to nucleus composition properties
is correct) may suggest another scenario. Models examining the thermal
evolution of the relatively large Kuiper Belt object show that due to radiogenic
heating, the most internal layers reach high temperatures \citep{Pri08,Sar09};
released gases migrate towards colder regions where they refreeze. The more
volatile ices refreeze closer to the cold surface than the less volatile, so
that the pristine dust/ice mixture becomes enriched in volatile ices such as
CO and N$_2$ \citep{Pri08,Sar09}. Dynamical studies of the transneptunian
population argue for a rich collisional history in the Kuiper Belt
\citep[][and references therein]{Mor15}. In the second scenario, comet
C/2016~R2 would be a fragment of the disruptive collision of a large Kuiper
Belt object, with properties representative of volatile-enriched layers.     
 
Comet C/2016~R2 is representative of a family of comets that we
observe only rarely each century. Besides C/1908~R1 (Morehouse) and C/1961~R1
(Humason) \citep{Bau11,Gre62}, other candidates are comets
29P/Schwassmann-Wachmann 1 and C/2002~VQ$_{94}$ (LINEAR) \citep{Kor08} which both
showed optical spectra dominated by strong emissions of CO$^+$ and N$_2^+$,
characteristics of abundant CO and N$_2$ production and high N$_2$/CO ratio.
The diversity of the dust/gas ratios seen in these comets (as judged from the
Af$\rho$ values, which indicate that 29P and C/2002~VQ$_{94}$ are dust-rich in
comparison to the other comets in the sample) may favour the second scenario
in which these comets are collisional fragments of differentiated
transneptunian objects.




\begin{acknowledgements} IRAM observations were conducted under the target of opportunity 
proposal D06-17 and we gratefully acknowledge the support from the IRAM
director for awarding us discretionary time and the IRAM staff for their
support and for scheduling the observations at short notice.
Observations of comet C/2014 S2 (PanSTARRS) were in part made during the eighth
IRAM summer school in September 2015, with contributions from P. Gratier,
E. Garcia  Garcia, A. Khudchenko, I. Kushniruk, T. Michiyama, A. Petriella,
F. Ruppin, Y. Shoham and P. Torne.
The data were reduced and analysed thanks to the use of the GILDAS,
class software (http://www.iram.fr/IRAMFR/GILDAS).
The Nan\c{c}ay Radio Observatory is the Unit\'e scientifique de
Nan\c{c}ay of the Observatoire de Paris, associated as Unit\'e de service
et de recherche (USR) No.704 to the French Centre national de la
recherche scientifique (CNRS). The Nan\c{c}ay Observatory also gratefully
acknowledges the financial support of the Conseil r\'egional of
the R\'egion Centre in France.
This research has been supported by the Programme national de 
plan\'etologie de l'Institut des sciences de l'univers (INSU).

\end{acknowledgements}


\end{document}